# Topological superconductivity in metallic nanowires fabricated with a Scanning Tunnelling Microscope

J G Rodrigo[1], V Crespo[1], H Suderow[1], S Vieira[1], F Guinea[2]

[1] Laboratorio de Bajas Temperaturas,
Departamento de Física de la Materia Condensada,
Instituto de Ciencia de Materiales Nicolás Cabrera, Facultad de Ciencias,
Universidad Autónoma de Madrid, E-28049 Madrid, Spain

[2]Instituto de Ciencia de Materiales de Madrid, CSIC
Sor Juana Inés de la Cruz 3
E-28049 Madrid, Spain

**Abstract.** We report on several low temperature experiments supporting the presence of Majorana fermions in superconducting lead nanowires fabricated with a scanning tunneling microscope, STM. These nanowires are the connecting bridges between the STM tip and the sample resulting from indentation-retraction processes. We show here that by a controlled tuning of the nanowire region, in which superconductivity is confined by applied magnetic fields, the conductance curves obtained in these situations are indicative of topological superconductivity and Majorana fermions. The most prominent feature of this behaviour is the emergence of a zero bias peak in the conductance curves, superimposed on a background characteristic of the conductance between a normal metal and a superconductor in the Andreev regime. The zero bias peak emerges in some nanowires when a magnetic field larger that the lead bulk critical field is applied. This field drives one of the electrodes into the normal state while the other, the tip, remains superconducting on its apex. Meanwhile a topological superconducting state appears in the connecting nanowire of nanometric size.

# 1. Introduction

Since Kitaev's proposal [1] in 2001 that Majorana fermions could be found in condensed matter systems a lot of efforts, both theoretical and experimental, have been addressed to the establish the conditions and requirements to detect such elusive particles. It was realized that topological insulators were a platform for Majorana states[2]. Most of these activities involve the use of the superconducting state, because of the close similarity between a Majorana fermion (a particle that it is its own antiparticle) and the quasiparticles of a p-wave spinless superconducting condensate that could be found close to (or at) its ground energy state (i.e. Fermi level) under some specific conditions and requirements. The different possibilities, requisites and conditions to be fulfilled by a superconducting system in order to present Majorana fermions have been extensively discussed in the recent years [3-9]. One of the "easiest" ways to obtain such p-wave spinless superconducting condensate consists of using a standard s-wave superconductor, whose electronic bands are splitted depending on the spin polarization by means of a spin-orbit (SO) coupling, and an external magnetic field which opens a Zeeman gap and a region without spin degeneracy. The effective spinless regime, where the standard superconducting states becomes a topological one, will be strongly dependent on the values of the superconducting energy gap, the Zeeman gap, the spin-orbit coupling and the filling of the bands (i.e. Fermi level). The existence of well defined energy bands and levels will favour the fulfilment of the required conditions. Therefore, 1D or quasi 1D systems with a small number of quantum modes will be more suited to the emergence of Majorana fermions.

The experimental efforts towards the realization of these requirements have focused mainly in the use of quasi-1D semiconducting nanowires [10-12], which present a strong spin-orbit coupling, and where superconductivity is induced by proximity to a s-wave superconductor. Mourik et al. [11] have reported recently experiments on such a system, showing evidences of the detection of Majorana fermions in a semiconducting nanowire by means of tunnelling spectroscopy measurements.

In a recent publication [13] we have shown a different approach to the experimental realization of the above mentioned "recipe". Instead of using a semiconductor as the "source" of the spin-orbit coupling we use lead, a metallic s-wave superconductor below 7K, which presents a moderate spin-orbit coupling.

Indeed, there are several differences between the superconducting state developed in a semiconductor and the one present in a metal, mainly related to the characteristic energy and length scales of the superconducting condensate which are closely related to the Fermi wavelength in each type of material.

Another important aspect is related to the requirement of a small number of quantum electronic modes involved in the experimental object. In the case of semiconductors, due to its low electronic density of states, this is quite easily achieved by using quasi-1D nanowires, with diameter in the range of 100 nm and 1 micron long. Such nanowires may be gated in order to move the Fermi level so that the above mentioned requirements are achieved. However, if we use a metal, the condition of having a small number (of the order of one) of quantum modes, or channels, is only achieved if the diameter of the nanowire, where topological superconductivity will be induced, is of the order of a few atoms [14]. This small (atomic) dimension of the nanowire imply that the effective superconducting coherence length will be also strongly reduced. Random scattering at the boundaries imply that the elastic mean free path, $\ell$, will be of order of the sample width, which, for a small number of channels within the nanowire, cannot be larger than a few nanometers. Then, the superconducting coherence length, $\xi \approx \sqrt{\xi_0 \ell}$ will be significantly reduced. For $\xi_0 \approx 80$ nm, and , $\ell \approx 1$ nm we find $\xi \approx 9$ nm, which is probably comparable with the nanowire length.

We have addressed the realization of such atomic-scale nanowire by using a STM which allows to establish atomic scale contacts between metallic electrodes (usually called "tip" and "sample") [15-16]. By using the different capabilities and features of the STM system it is possible to control and modify this atomic scale contact, even detecting the addition of individual atoms to the contact [17]. As an example, it has been possible to create gold nanowires consisting in a chain of single atoms between the tip and sample gold electrodes, where the electric conductance involves a single quantum channel [18].

The low value of the spin orbit coupling in lead, compared to the value in some semiconducting materials, is an added difficulty in order to accomplish with the requirements to create a topological superconducting phase in the nanowire. Lead is a type-I superconductor, with a superconducting gap of 1.35 meV at zero field and zero temperature, and a rather low bulk critical field (75 mT at 300mK). However, the nanoscopic apex of the tip, and sharp elongated nanostructures, may remain superconducting well above the bulk critical field[16]. Thus, a magnetic field in the range of 100 mT will produce a Zeeman splitting 0.04–0.06 meV and, as the gap is expected to go smoothly to zero as the magnetic field increases, it is possible that a regime where the Zeeman coupling is larger than the superconducting gap is present in some of these lead nanowires. The position of the Fermi energy at the nanowire will depend on details of the electrostatic potential which, in turn, is determined by the geometry of the contact and the voltage distribution within it, which, in turn, depends on the bias voltage. Therefore, it is possible that some of such Pb nanowires present a Fermi energy level located within the Zeeman gap, due to random fluctuations in the electrostatic potential. It might be possible to tune the parameters of the device by changing the bias voltage, although a detailed analysis of this topic lies outside the scope of the present work.

Our experimental configuration, shown schematically in figure 1, presents another peculiarity that supposes a relevant difference compared to the above mentioned experiments involving semiconductors. The lead nanowire, which will become a topological superconductor under an external magnetic field, is created by indenting the STM Pb tip (which in the end is of nanoscopic dimensions) into a Pb surface (which can be considered "flat" compared to the tip).

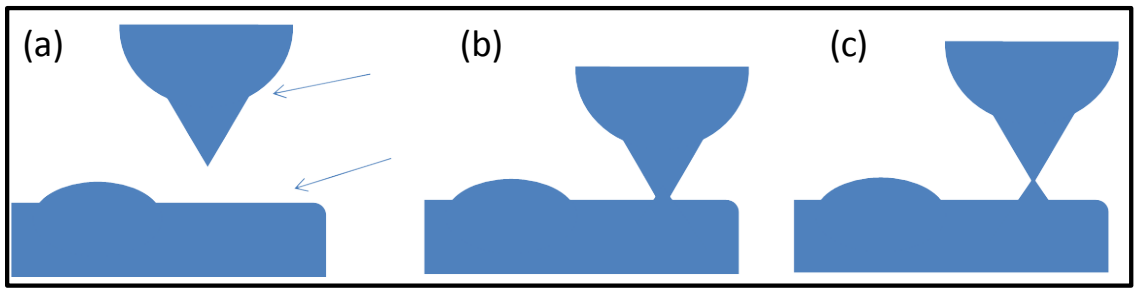

**Figure 1** Schematics of the experimental configuration. A STM tip, with an atomically sharp nanotip, is brought to contact onto a sample of the same material, lead in our case (a,b). The STM control system can be used to adjust the tunneling current, control the size of the contact established between tip and sample and eventually create a nanostructure between tip and sample by means of indentation-retraction cycles.

We have shown in previous works [16, 19-20] that sharp and elongated nanotips or nanoprotrusions on the sample surface resulting from the tip-sample indentations remain superconducting for magnetic fields much larger than the bulk critical field of lead, depending on their sharpness and dimensions compared to the bulk coherence length and penetration depth, while the larger bulk, flat or blunt parts of the Pb tip and sample have become normal.

Thus, we have a system where the topological superconducting Pb nanowire, which may present Majorana fermions, is in direct contact with (has emerged from) a bulk of the same material in normal state, and it is probed by a superconducting electrode (the nanotip) using and involving a very small number of quantum channels (typically between 3 and 10).

The probe, the nanotip, is in direct contact with the nanowire, therefore the spectroscopic measurments will take place in a transmission regime where Andreev reflections play a key role [21-22]. The shape and features of the spectroscopic curves (IV curves) in this Andreev regime are strongly dependent on the coupling and transmission probability of each individual quantum channel [23-25], thus providing a detailed quantification of the number of quantum channels involved in the conduction process (and at the nanowire), and their individual transmission.

In the following sections we present our recent experiments on these topological nanowires. We discuss on the experimental evidences of the emergence of a topological superconducting phase at the nanowire/nanocontact region; the comparison of this type of results with those that could be considered "expected" or "normal" for standard S-S or S-N situations; the observability of such "unexpected" results, which we assign to the presence of Majorana fermions, depending on the number of quantum channels involved and their transmissions; and the analysis of the zero bias current detected in the different superconducting regimes of the nanowire. The evolution of this zero bias current may be of key importance in the study and detection of Majorana fermions, as it would be a direct indication of the splitting of Cooper pairs injected from the superconducting nanotip into pairs of Majorana bound states in a topological superconducting nanowire with extremely small superconducting gap.

## 2. Experiments

### 2.1 Sample fabrication

The experiments are performed at 0.3 K, with the STM installed in a $^3$He cryostat equipped with a superconducting solenoid. The STM is used to produce indentations, in the range of a few tens of nanometers, of a Pb tip on a Pb sample, in order to fabricate sharp elongated nanotips and nano-protrusions on the sample surface. Along these indentation processes we record the variation of the current across the contact as a function of the relative displacement between tip and sample, for a fixed bias voltage (typically in the range of 10 mV). The analysis of the current vs. displacement curves allows to extract information about the sharpness and dimensions of the nanostructures resulting of the indentation process [15]. When the contact is broken (by receding the tip) it is possible to scan and visualize the part of the nanostructure remaining on the sample surface (figure 2).

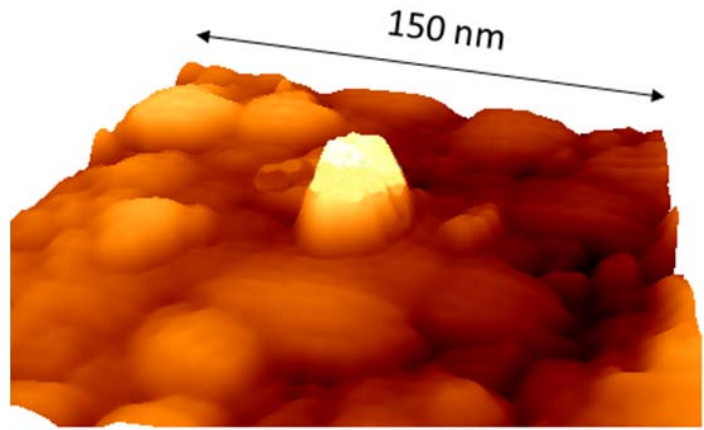

**Figure 2.** STM topographic image of a Pb surface where a nanometer size protusion, created after a tip-sample indentation process, is clearly visible. The nanostructure is 35 nm wide at its base, and 25 nm high. The actual lateral dimensions of the nanostructure will be smaller, as it is scanned by an equivalent protusion at the tip apex. The image was taken at 0.3K, for a tunnelling resistance of 10 MΩ, and voltage bias of 10 mV.

Once a sharp Pb nanotip has been created, we move to a "flat" region of the sample (far from the nanoprotusion) and we proceed with the spectroscopic characterization of the nanotip. Making use of the STM feedback control loop we go from tunnelling conditions towards atomic scale contact between tip and sample. This process is followed in detail by recording the current vs. displacement variations and the acquisition of current vs. voltage curves along the process. The IV (and conductance, dI/dV vs V) curves evolve from the usual gapped structure in tunneling to a much richer structure (the subharmonic gap structure) as the contact is established, due to the increasing contribution of Andreev reflection processes.

As mentioned above, this rich structure in the spectroscopic curves allows to determine precisely the number of quantum channels and their individual coupling or transmission (figure 3). A zero bias current is easily observed in the curves, signature of transmission of Cooper pairs between the electrodes (dc Josephson effect). Typically, voltage bias is ramped between ±10 mV, and the conductance is obtained by numerical derivative of the IV data.

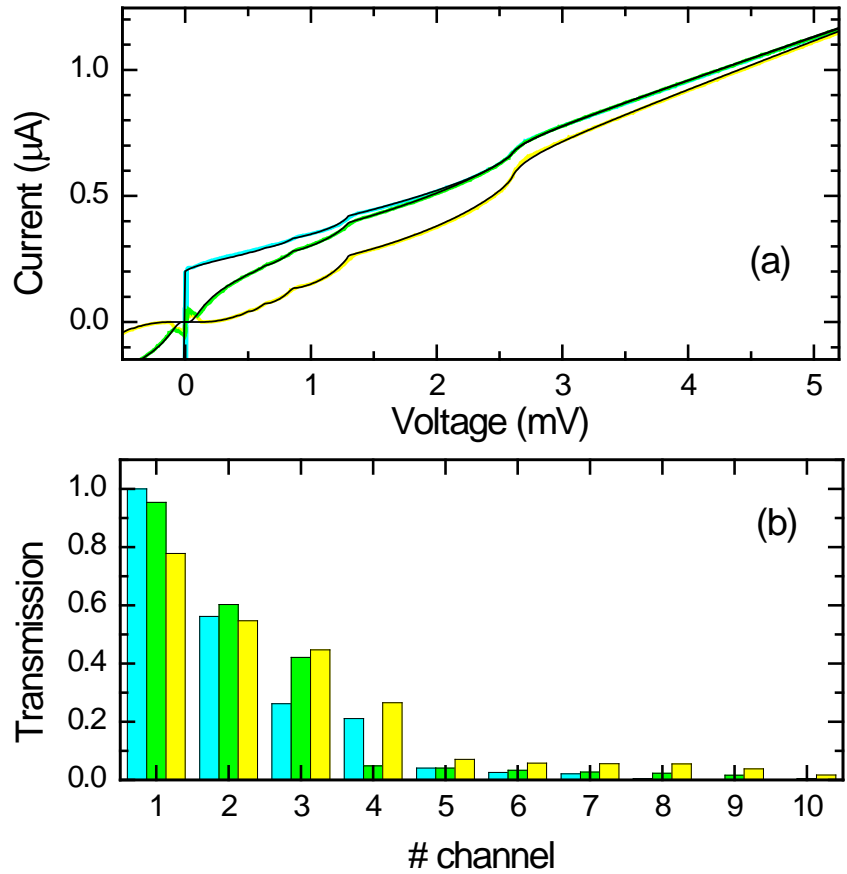

**Figure 3.** (a) Experimental IV curves (colour lines), and calculated fittings considering quantum channels and Andreev reflections (black lines). In (b) we show the corresponding quantum channels and transmissions. The curves were taken after small atomic rearrangements of a nanocontact, at 350 mK and zero magnetic field.

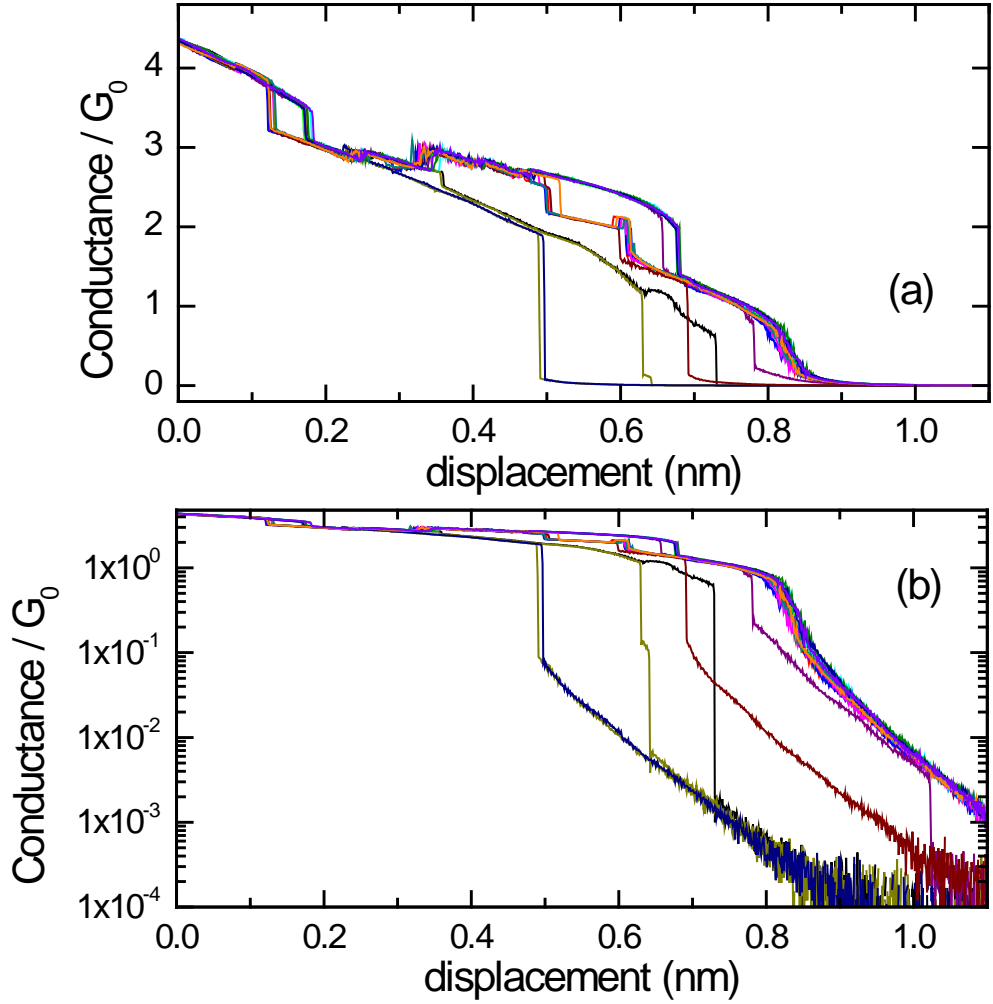

**Figure 4.** (a) Experimental curves obtained in a series of 8 consecutive indentation-elongation cylces of a Pb tip into a Pb sample. From cycle to cycle, it can be seen that the nanocontact (i.e. the nanowire) between tip and sample may be elongated in different ways before breaking it and entering in tunnelling regime (see the exponential variation of the conductance vs. electrode displacement in the logarithmic plot (b))

Once the atomic scale contact is created, we can perform small indentation cycles, with nominal displacements smaller than 10 nm. This leads, in each cycle, to atomic scale rearrangements in the nanocontact region (figure 4). Eventually, these mechanical processes result in a nanometric length wire, a nanowire, whose cross section is made of a small number of atoms (as extracted from the quantum channel analysis of the conductance curves). These nanowires are expected to present a topological superconducting state, and eventually Majorana fermions, when a magnetic field is applied.

In these experiments, different nanostructures, with conductance values at the constriction ranging from $2\,G_0$ to $50\,G_0$ ($G_0 = 2e^2/h$ is the conductance quantum) were created, and we have analyzed the evolution of their spectroscopic characteristics vs. magnetic field in order to investigate the presence of Majorana fermions.

## 2.2 Sample characterization

As shown in figure 5, different types of conductance curves (i.e., dI/dV vs V) can be obtained depending on the geometry of the electrodes at the nanoscale close to the contact, and the values of the external magnetic field [20, 26]. Figure 5(a) shows the typical result obtained at zero field, and for H<Hc bulk. The full subharmonic gap structure due to multiple Andreev reflections, and a sharp zero bias peak due to Josephson current are obtained, being the exact features depending on the transmissions of the different quantum channels involved in the process. The other panels show the more representative types of conductance curves that can be obtained for H>Hc bulk. We also show a sketch of the geometry of the nanostructure leading to each observed result. If the contact is established between similar sharp protusions in tip and sample, both remain superconducting at H>Hc, showing results similar to the ones at zero field. If the sample side of the nanostructure is more rounded or blunt than the nanotip, S-S features are still present, but very rounded due to pair-breaking effects at the sample, now more sensitive to the effect of the applied magnetic field (figure 5(b)). When the nanotip contacts a flat region of the sample we obtain results like the one in figure 5(c), corresponding to a situation involving NS Andreev reflections, where only the nanotip remains superconducting, and all the sample is in normal state. Finally, in figure 5(d) we show a type of conductance curve that can be obtained after modifying situations like the ones shown in figure 5(c) by means of indentation-retraction cycles. The new feature is a zero bias peak in the conductance, corresponding to a finite zero bias current, which is superimposed on the NS Andreev-like conductance for finite bias. We consider that, in this case, a nanowire is created between tip and sample.

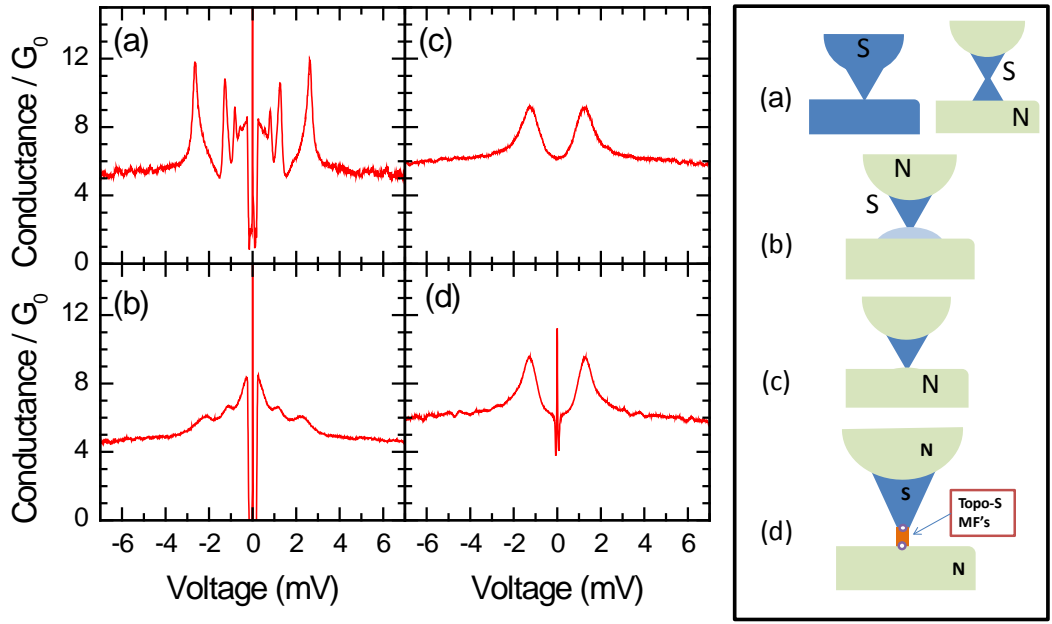

**Figure 5.** Different types of conductance curves that can be obtained in the studied nanostructures and nanocontacts. Left panel shows the different types of conductance curves: (a) S-S type: for H<Hc, and for H>Hc if there are sharp protusions at nanotip and sample that remain in superconducting state; (b) S-S' type: for H>Hc, corresponding to a sharp nanotip and a blunt nanostructure at the sample, which presents superconductivity with strong pair breaking; (c) S-N type: for H>Hc, the sharp nanotip remains

superconducting, and the flat sample is in normal state; (d) in this type, a zero bias peak appears for H>Hc, and we consider that a nanowire is created between the sharp S nanotip and the N sample. In the right panel we sketch the geometries corresponding to the different types of curves.

In the following, we will describe the evolution with magnetic field of the conductance curves in the different nanostructures presented above.

## 3. Experimental results and discussion

In fig.6(a) we show the evolution of the conductance curves vs. magnetic field in a situation where no "anomalous" or topological superconducting state is expected to happen. The conductance of this nanostructure is 6 $G_0$. The atomically sharp nanotip is contacting a "flat" region of the sample and no indentation-elongation cycles were performed. In this situation, as we increase the external magnetic field, once the bulk critical field of lead is crossed, we observe a sharp transition from SS spectroscopic curves (showing a very rich subharmonic gap structure and Josephson current) below Hc, to NS like curves, with no indication of zero bias current. Obviously, for H>Hc only the nanotip remains superconducting, and the bulk parts of both electrodes has become normal.

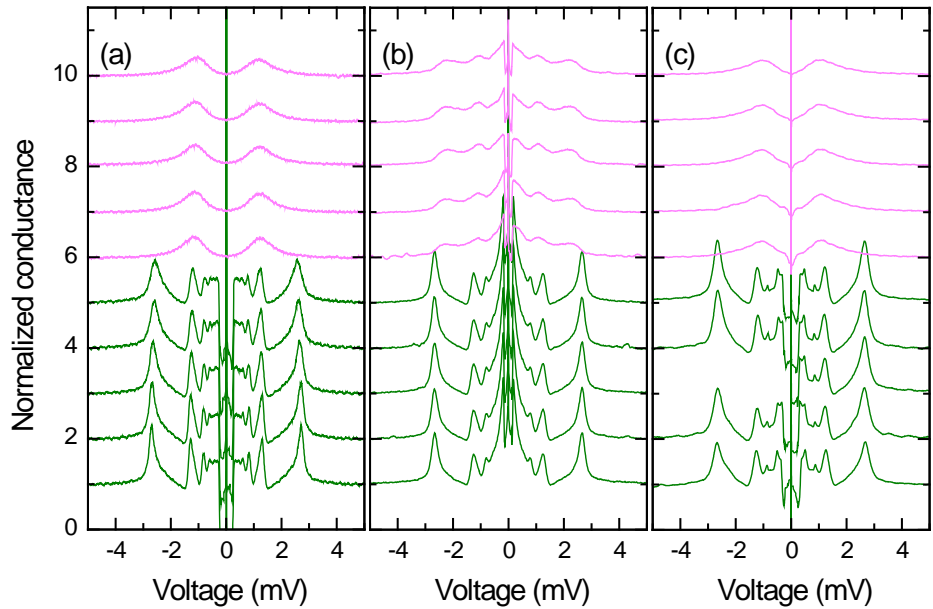

**Figure 6.** Examples of typical evolution of the spectroscopic conductance curves of different nanowires as the applied magnetic field is varied across its bulk critical value (Hc = 75 mT). The panels show curves obtained at magnetic fields from 30 mT (bottom) to 120 mT (top), in 10 mT increments. Curves in green correspond to H<Hc, while magenta is used for H>Hc. In (a) there is a clear and sharp transition from well defined curves showing standard SS Multiple Andreev reflections features below Hc, to NS Andreev curves for H>Hc, corresponding to a sharp tip contacting a flat region on the sample. In (b), the tip was located on a bump on the surface, resulting in rounded spectra for H>Hc, due to the bump being still in superconducting state, but under strong pair breaking effects due to the field. Panel (c) shows the conductance curves that we consider due to the presence of

Majorana states at the constriction: for H>Hc a zero bias peak appears, superimposed on a conductance background corresponding to an apparent NS situation, similar to those in (a). Curves are normalized at their high voltage value, and shifted vertically for clarity.

If the contact is established with the nanotip located onto a "bumpy" spot of the sample, this region may remain superconducting above Hc, but showing strong pair breaking effects (rounding and shift of the peaks corresponding to the subharmonic gap structure, accompanied by a progressive decrease of the signal at zero bias), and eventually becomes normal. An example of this case is shown in figure 6(b), corresponding to a nanostructure with conductance 4 $G_0$, and it is also considered as an expected one [20].

As we showed in a recent work [13], some nanostructures present a peculiar behaviour at magnetic fields above Hc. The corresponding spectroscopic curves show, for V≠0, clear and well defined NS features, which can be fitted in terms of quantum channels and NS Andreev reflections (indicating that only one electrode remains superconducting, the nanotip), while a finite current at zero bias, like the one corresponding to the dc Josephson effect in the S-S case, can still be clearly detected (figure 6(c)). We have observed that, under H>Hc, the zero bias feature may disappear, and eventually reappear, after performing small indentation-elongation cycles.

We consider this behaviour, the emergence of a zero peak, superimposed onto a standard NS Andreev-like quasiparticle curve, as a signature of Majorana fermions in the nanowire that has been created between tip and sample. As discussed earlier, we describe the wire as dirty superconductor, with the mean free path, $\ell$, limited by the width of the wire. The effective coherence length within the wire is reduced to $\xi \approx 10$ nm. Hence, the two ends of a sufficiently long wire are decoupled, and the Andreev states pinned at them can be considered independently, see figure 7, left, where we also sketch the process needed to obtain zero bias current between a superconductor an a normal electrode, mediated by the presence of a topological superconductor with Majorana fermions. In the standard N-S case, when the topological superconducting nanowire is not present, no current can be obtained at zero bias, because of the presence of the energy gap in the superconductor, and the requirement of finite difference between the Fermi levels in N and S in order to Andreev reflection processes to be allowed.

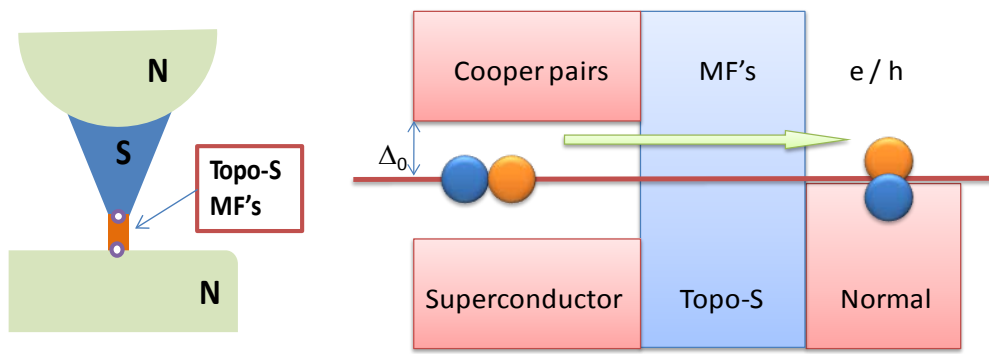

**Figure 7.** (a) Schematics of the experimental system, indicating the different basic components and their state under H>Hc : the nanotip, superconducting; the nanowire, topological superconducting with Majorana fermions; and the sample, in normal state. (b)

Schematics of the transport process at V=0, leading to finite current, mediated by the Majorana fermions.

Next, we investigate the experimental conditions that allow us the observation of the phenomena that we ascribe to the presence of a topological superconducting nanowire with Majorana fermions. As it was mentioned in the introduction, it is possible to extract the number of quantum channels and their transmissions involved in the transmission processes leading to the different spectroscopic curves. This can be done both in the low field regime (H<Hc bulk, SS characteristics) and in the topological situation (H>Hc bulk, NS quasiparticle characteristics). In figure 8 we show calculated one-channel conductance curves corresponding to different values of the transmission probability. The conductance curve corresponding to a given experimental situation will be the result of the addition of the conductance from different channels with different transmissions, thus resulting in a unique type and shape of the conductance curve in each case.

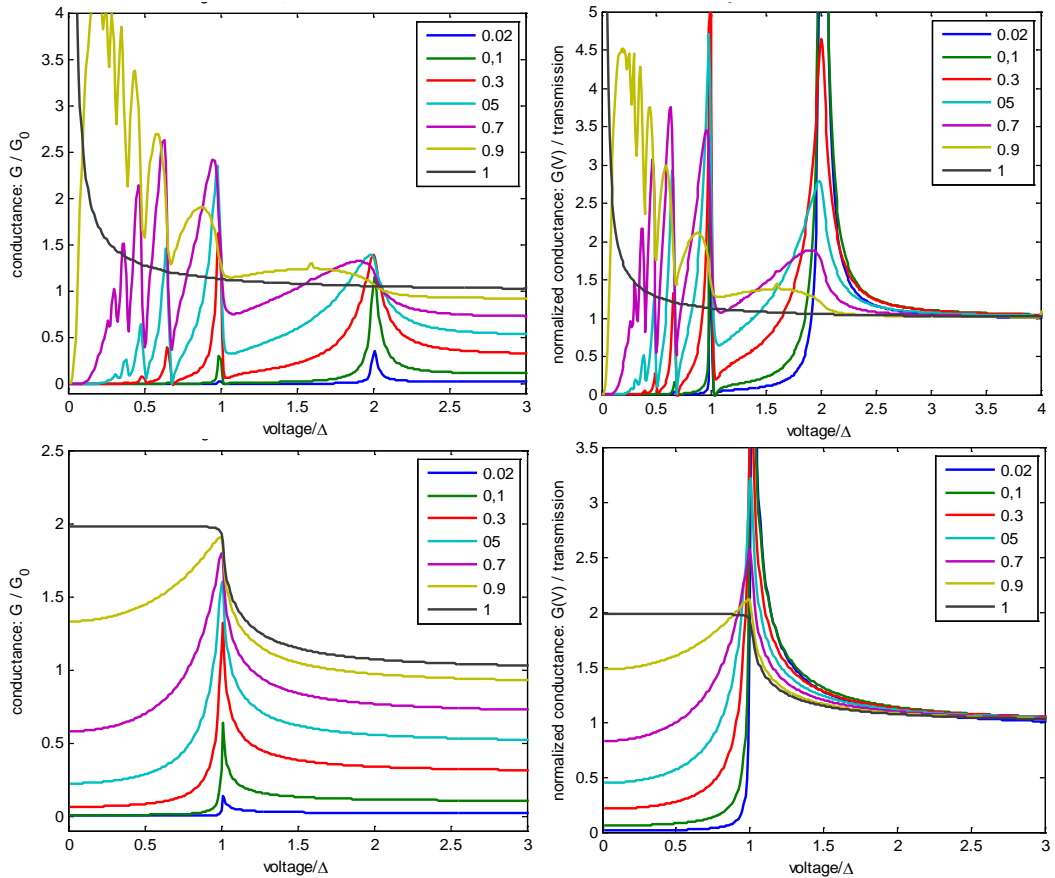

**Figure 8.** Calculated conductance curves corresponding to different values of the transmission probability of a single quantum channel. We show the SS situation (top), and the NS case (bottom). The curves are calculated for a single channel, with transmissions from 0.02 to 1. The NS case was calculated introducing a small value for the magnetic induced pair breaking, $\Gamma=0.005\ \Delta$.

The analysis of tens of nanowires showed that, for similar values of the total conductance, typically in the range of 3-10 $G_0$, only those that presented high transparency channels (with transmission probability very close to one) showed a well defined zero bias current together with a NS quasiparticle Andreev regime, independently of how was the channel distribution in the SS low field regime. In figure 9 we show two examples illustrating the different situations.

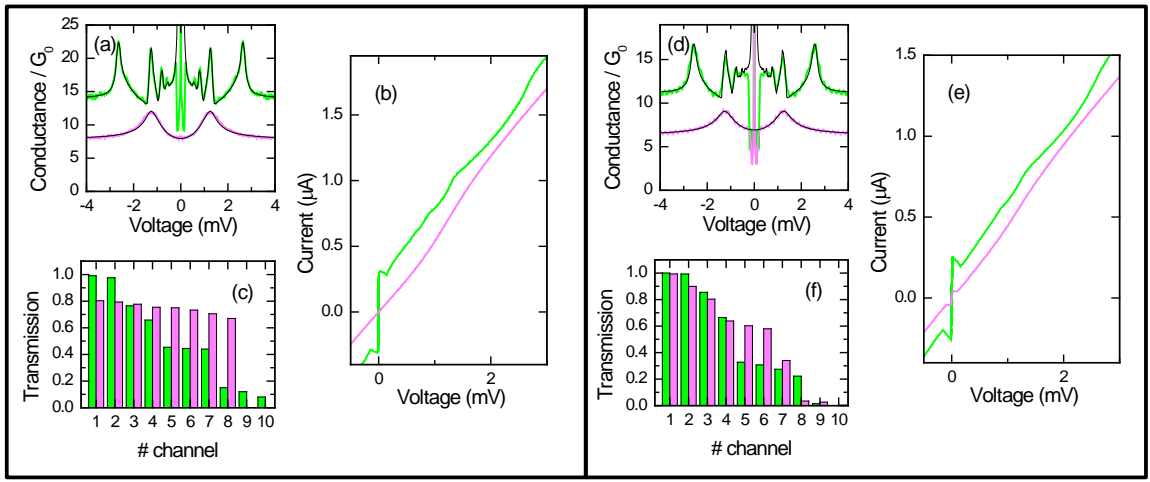

**Figure 9.** Conductance curves, and their corresponding fittings in terms of quantum channels and Andreev reflections, for two different nanocontacts, showing curves taken below Hc (green) and above Hc (magenta). Left panel shows a "standard" SS to NS transition in a contact with conductance 8 $G_0$, while for the contact shown in the right panel, with conductance 7 $G_0$, a sharp peak at zero bias appears superimposed on an NS conductance for H>Hc, corresponding to a finite zero bias current. These two types of behaviour could be achieved and tuned by modifications of the atomic arrangement in the nanowire by means of indentation-elongation cycles in the range of 1nm. Panels (a) and (d) show the conductance curves. Panels (b) and (e) show a zoom of the region of the IV curves close to zero bias. Panels (c) and (f) show the distributions of transmissions of the quantum channels used to generate the corresponding calculated fitting curves, shown in black in panels (a) and (d). (Conductance curves in the low field regime (green) are shifted vertically for clarity).

The value of the zero bias current in the topological regime (magenta curves, in figure 9(d-f)) was in these cases always in the range of 1/4 to 1/5 of the value obtained at zero field (or for H<Hc bulk). This observation can be taken as a proof to rule out the possibility that this zero bias current in the topological regime is just Josephson current between the superconducting tip (with a gap value equal to the one at zero field, $\Delta_0$) and a superconducting region of the sample with a very reduced gap. In the conductance curve no signature of a second gap is detected, anyway we could say that a high limit to this second gap, $\Delta_2$, would be 2% of $\Delta_0$, while the gap at the tip, $\Delta_1$, equals $\Delta_0$. Josephson current in such an $S_1$-$S_2$ situation would be of the order of $\Delta_1\Delta_2/(\Delta_1 + \Delta_2)$, leading to a value of the order of $\Delta_0/50$, a quantity 25 times smaller than $\Delta_0/2$, the value corresponding to identical gaps. Note than in our experiments, the zero bias current is reduced just a factor 4 or 5, not 25, with respect to the case when both electrodes are in superconducting state.

The relevance of the presence of high transmission channels in order to observe the zero bias feature (i.e. Majorana fermions) has been addressed by studying nanowires whose total conductance is slightly modified in a controlled way. With the nanowire in the topological regime (H>Hc bulk) we vary the conductance a value equivalent to $1\,G_0$ or less. The example in figure 10(a) shows that the curve with larger total conductance presents a slightly smaller zero bias peak. This behavior is very similar to the one observed in S-S nanocontacts [27], where it was seen that the value of the zero bias current depends mainly on the individual transmissions of the quantum channels, and not on the total conductance of the nanocontact. These observations are compatible with the analysis in ref. [28], where the robustness of the Majorana signatures in high transparency situations was studied.

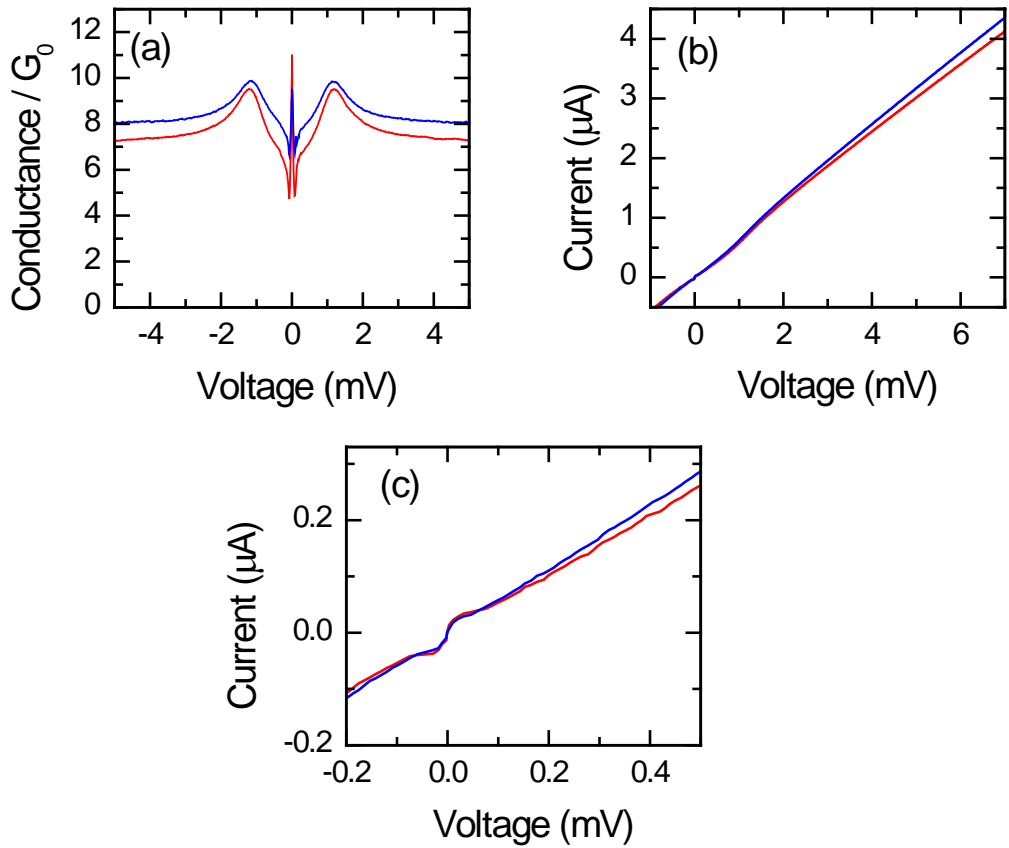

**Figure 10.** Example of a situation where a slightly increased conductance of the nanowire (blue curves, (a)), it is not accompanied by an increase of the zero bias current, as can be seen in (b) and the zoomed zero bias region (c).

Finally, we show an example of the evolution of the zero bias current with temperature. We wish to note that different nanowires could present slightly different behaviors, but the overall behavior is the one depicted in figure 11. We prepare a nanostructure with a conductance value of 5 $G_0$, and apply a magnetic field of 200 mT. As temperature increases, the Andreev-gap related features in the conductance curves are rounded and move towards zero bias (i.e. the gap at the nanotip decreases as temperature increases), disappearing at about 6K in this case, while the signature corresponding to zero bias current was observed to disappear at a much lower T, between 1.5K and 2K. This is an indication of a critical temperature for the topological superconducting nanowire lower, between 25% and 33%, than the one of the "parent" superconducting material.

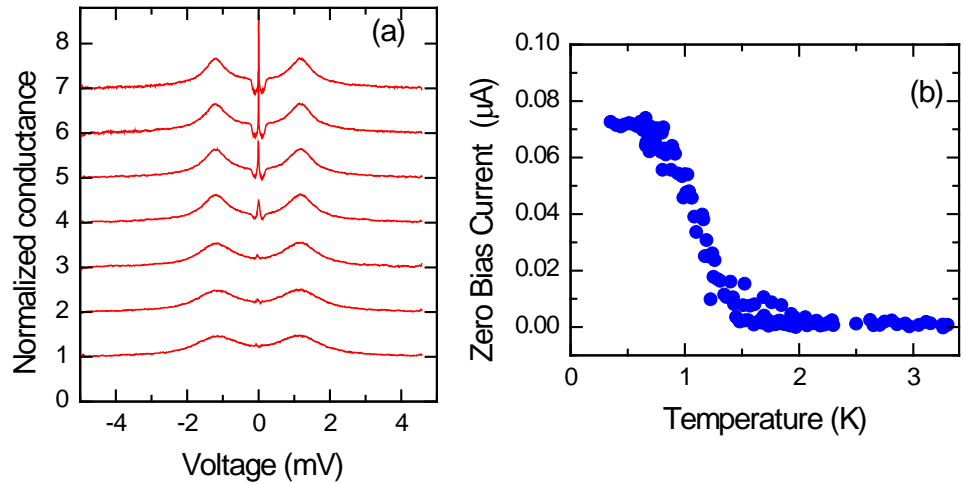

**Figure 11.** Evolution of the zero bias feature with temperature in the topological superconducting regime of the nanowire. (a) Selected conductance curves (normalized and shifted vertically) obtained at 200 mT for different temperatures (Temperature increases from top to bottom, from 0.4K to 1.9K in increments of 0.25K ). The gap-like feature due to Andreev reflections was observed to disappear at 6K. (b) Variation of the zero bias current with temperature.

In a recent work, Potter and Lee [29] proposed a system where surface states in a metallic gold nanowire (in contact with a s-wave superconductor) may fulfil the requirements to present topological superconductivity and Majorana fermions. They suggest that Majorana fermions, arising from surface states in the metallic nanowire, could be detected by a STM tip. The nanowire would have dimensions similar to the ones used in semiconducting wires (in the range of 100 nm wide and 1000 nm long), and should be grown in the (111) orientation. In our experimental realization, a metallic conducting nanowire is also considered to develop topological superconductivity and Majorana fermions. However, the nanowire itself is made of a superconducting material, lead, and brought to atomic scale in order to carry a few quantum channels. Moreover, in our case, the superconducting gap at the nanowire is reduced (or almost destroyed) by the external magnetic field, and it is not proximity-induced by a "back-up" s-wave superconductor. We consider that these differences, shown schematically in figure 12, should be considered in further theoretical works dealing on the presence and detection of Majorana fermions in topological superconducting metallic nanostructures.

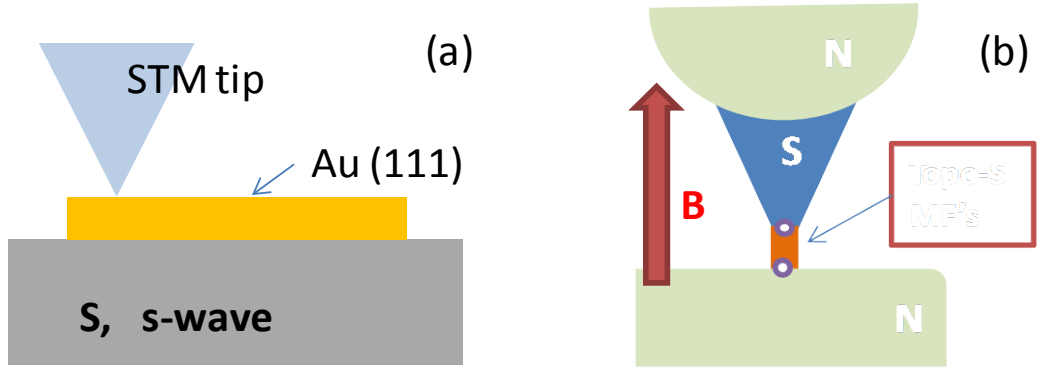

**Figure 12.** Sketch of the experimental systems proposed by Potter and Lee involving a metallic nanowire (a), and the one proposed in this work (b).

## 4. Conclusions

We have reported on a study on metallic superconducting nanostructures and nanowires which, under given values of the external magnetic field and number and coupling of the quantum channels at the nanowire, may lead to the observation of features that can be ascribed to the presence of Majorana fermions when a topological superconducting state is induced in the nanowire. We have shown that the channel number and transparency are key elements in the detection of Majorana fermions in these systems, which appears as a zero bias current superimposed onto a standard NS quasiparticle curve in the Andreev reflection regime. Application, using the STM capabilities, of small elongation-retraction cycles of the order of 1 nm to the nanowire, may lead to the disappearance or appearance of the zero bias signature. When the conductance curves are analyzed we obtain that the presence of a few high transparency channels are generally associated to the presence of the zero bias peak.

We consider that the metallic systems presented in this work can be an alternative to other systems, involving mainly semiconductors, currently under study in the search for Majorana fermions in condensed matter physics. An open question, to be considered in future work, is the fine tuning of the parameters which characterize the devices. This can be achieved by small changes in the structure once the basic features have been fixed, and by control of the region where the voltage drops, which, in turn, depends on the bias voltage.

**Acknowledgments**

The Laboratorio de Bajas Temperaturas is associated to the ICMM of the CSIC. This work was supported by the Spanish MINECO (Consolider Ingenio Molecular Nanoscience CSD2007-00010 program, FIS2011-23488), by the Comunidad de Madrid through program Nanobiomagnet. F. G. acknowledges funding from grants FIS2008-00124, FIS2011-23713 and from the ERC Advanced Grants program, contract 290846. *This research was supported in part by the National Science Foundation under Grant No. NSF PHY11-25915*. F. G. acknowledges useful conversations with E. Prada, P. San Jose, and B. Trauzettel.